\documentclass[journal=jacsat,manuscript=article]{achemso}

\usepackage[version=4]{mhchem} 
\usepackage{siunitx}
\usepackage{multirow, makecell}
\usepackage{tabularx}
\usepackage[table]{xcolor}

\author{KeYuan Ma}
\affiliation[University of Zurich]
{Department of Chemistry, University of Zurich, CH-8057 Z\"urich, Switzerland}

\author{Robin Lef\`evre}
\affiliation[University of Zurich]
{Department of Chemistry, University of Zurich, CH-8057 Z\"urich, Switzerland}

\author{Karolina Gornicka}
\affiliation [Gdansk University of Technology]
{Faculty of Applied Physics and Mathematics, Gdansk University of Technology, Gdansk 80-233, Poland}
\altaffiliation{Advanced Materials Centre, Gdansk University of Technology, Gdansk 80–233, Poland}

\author{Harald O. Jeschke}
\affiliation[Okayama University]
{Research Institute for Interdisciplinary Science, Okayama University, Okayama 700-8530, Japan}

\author{Xiaofu Zhang}
\affiliation[University of Zurich]
{Department of Physics, University of Zurich, CH-8057 Z\"urich, Switzerland}
\altaffiliation{State Key Laboratory of
Functional Materials for Informatics, Shanghai Institute of Microsystem and Information Technology, Chinese Academy of Sciences (CAS), Shanghai 200050, China}

\author{Zurab Guguchia}
\affiliation[Paul Scherrer Institute]
{Laboratory for Muon Spin Spectroscopy, Paul Scherrer Institute, CH-5232 Villigen PSI, Switzerland}

\author{Tomasz Klimczuk}
\affiliation [Gdansk University of Technology]
{Faculty of Applied Physics and Mathematics, Gdansk University of Technology, Gdansk 80-233, Poland}
\altaffiliation{Advanced Materials Centre, Gdansk University of Technology, Gdansk 80–233, Poland}

\author{Fabian O. von Rohr}
\affiliation[University of Zurich]
{Department of Chemistry, University of Zurich, CH-8057 Z\"urich, Switzerland}

\title{Group-9 Transition Metal Suboxides Adopting the Filled-\ce{Ti2Ni} Structure: A Class of Superconductors Exhibiting Exceptionally High Upper Critical Fields} 

\keywords{Suboxide Superconductors, $\eta$-carbide-type materials, superconducting magnets}

\begin{document}
\begin{tocentry}

\includegraphics[]{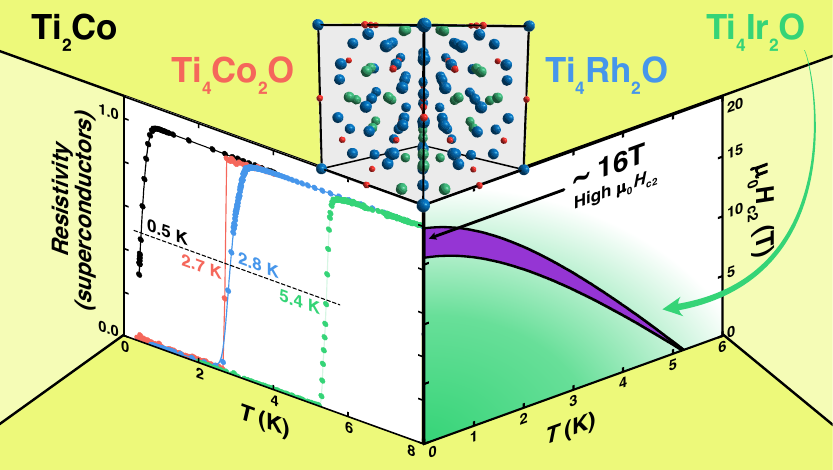}

\end{tocentry}

\begin{abstract}
The \ce{Ti2Ni} and the related $\eta$-carbide structure are known to exhibit various intriguing physical properties. The \ce{Ti2Ni} structure with the cubic space group $Fd\bar{3}m$ is surprisingly complex, consisting of a unit cell with 96 metal atoms. The related $\eta$-carbide compounds correspond to a filled version of the \ce{Ti2Ni} structure. Here, we report on the structure and superconductivity in the $\eta$-carbide type suboxides \ce{Ti4M2O} with M = Co, Rh, Ir. We have successfully synthesized all three compounds in single phase form. We find all three compounds to be type-II bulk superconductors with transition temperatures of $T_{\rm c}$ = 2.7, 2.8, and 5.4 K, and with normalized specific heat jumps of $\Delta C/\gamma T_{\rm c}$ = 1.65, 1.28, and 1.80 for \ce{Ti4Co2O}, \ce{Ti4Rh2O}, and \ce{Ti4Ir2O}, respectively. We find that all three superconductors, exhibit high upper-critical fields. Particularly noteworthy is in this regard \ce{Ti4Ir2O} with an upper critical field of $\mu_0 H_{\rm c2}{\rm (0)}$ =~16.06~T, which exceeds by far the weak-coupling Pauli limit -- widely consider as the maximal upper critical field -- of $\mu_0 H_{\rm Pauli}$~=~9.86~T. The role of the void filling light atom X has so far been uncertain for the overall physical properties of these materials. Herein, we have successfully grown single crystals of \ce{Ti2Co}. In contrast to the metallic $\eta$-carbide type suboxides \ce{Ti4M2O}, we find that \ce{Ti2Co} displays a semimetallic behavior down to 0.75 K. Below 0.75 K we observe a broad decrease in the resistivity, which can most likely be attributed to an onset of a superconducting transition at lower temperatures. Hence, the octahedral void-filling oxygen plays a crucial role for the overall physical properties, even though its effect on the crystal structure is small. Our results indicate that the design of new superconductors by incorporation of electron-acceptor atoms may in the \ce{Ti2Ni}-type structures and other materials with crystallographic void position be a promising future approach. The remarkably high upper critical fields, in this family of compounds, may furthermore spark significant future interest.
   
\end{abstract}


\section{Introduction}
The discovery of new superconductors and improved superconducting properties remains a long-standing challenge in solid-state and materials chemistry\cite{canfield2011still}. While there is a wide range of superconducting applications that would strongly benefit from new materials, state-of-the-art computational methods are struggling to predict the highly complex collective electronic state of superconductors.\cite{hutcheon2020predicting,yang2021testing} Hence, for the search of new and improved superconducting materials one has to rely on utilizing chemical and physical design principles, such as looking for new materials with structural features from known superconductors and electron counting. \cite{verchenko2020family,fujioka2016discovery,Ca3Ir4Ge4,stolze2018sc,rendenbach2021electrochemical,he2021superconductivity} One chemical approach to this has been the exploration of materials in which the electronic properties can be tuned precisely by the addition of dopants between layers or into voids positions.\cite{weller2005superconductivity,burrard2013enhancement,krzton2012synthesis,haldolaarachchige2014superconducting} For this purpose, electron-donor atoms, or cations, are most commonly incorporated into the structure, as e.g. in \ce{Cu_xTiSe2}, where superconductivity with a critical temperature of $T_{\rm c} \approx$  4 K was discovered via the intercalation of copper between the \ce{TiSe2} layers.\cite{morosan2006superconductivity} Less commonly, electron-acceptor atoms, or anions, are doped into structural void positions to achieve superconductivity \cite{zhang2017electride,kitagawa2020superconductivity}. A recent example for the enhancement of superconductivity in such a material, \ce{Nb5Ir3O} with a critical temperature of $T_{\rm c} \approx$  10.5 K \cite{zhang2017electride}. Here, oxygen \ce{O^{2-}} anions are occupying trigonal anti-prismatic channels in the \ce{Mn5Si3}-type host structure. 

All type-II superconductors share two fundamental attributes, (i) the critical temperature $T_{\rm c}$ below which superconducting state occurs and (ii) the upper critical field $H_{\rm c2}(T)$ above which superconductivity is fully suppressed \cite{Tinkham2004,gornicka2021nbir2b2}. The second parameter is crucial not only because it defines the potential practical applications of the superconductor, but also it contains fundamental information about the Cooper pair breaking process from the superconducting to the normal state. The maximal $H_{\rm c2}(0)$ for conventional superconductors is given by the paramagnetic pair breaking effect, which is commonly known as the Pauli paramagnetic limit $H_{\rm Pauli}$ \cite{Tinkham2004}. This value is given by $\mu_0 H_{\rm Pauli} = \frac{\Delta_0}{\sqrt{2} \mu_{\rm B}}\approx 1.86{\rm [T/K]} \cdot T_{\rm c}$ with $\mu_{\rm B}$ being the Bohr magneton and $\Delta_0$ being the superconducting gap in the weak-coupling Bardeen–Cooper–Schrieffer (BCS) theory. There are few unconventional superconductors (e.g. highly anisotropic superconductors\cite{tanaka2020superconducting,lu2014superconductivity,falson2020type}, heavy-fermionic superconductors\cite{bauer2004heavy,manago2017absence} and spin-triplet superconductor \ce{UTe2}\cite{aoki2019unconventional}) with high values of $H_{\rm c2}(0)$ that are exceeding their weak-coupling Pauli limit $H_{\rm Pauli}$. A variety of possible reasons for the violation of the weak-coupling Pauli paramagnetic limit have been discussed, including large spin-orbit coupling of the electrons or unconventional superconducting pairing mechanism \cite{carnicom2018tarh2b2,lu2014superconductivity, Fuchs2008,falson2020type}. Generally, a violation of the weak-coupling Pauli paramagnetic limit is widely considered to be a strong indicator for unconventional superconductivity \cite{gornicka2021nbir2b2, mercure2012upper,cao2021pauli}. 

$\eta$-Carbide-type compounds are a large family of compounds with more than 120 known members. $\eta$-Carbide-type compounds exist in two chemical stoichiometries, with the general formula \ce{A4B2X} or \ce{A3B3X}. Here A and B are transition metals, and X is a light element -- either carbon, nitrogen, or oxygen \cite{kuo1953formation, ma2019superconductivity}. The $\eta$-carbide structure corresponds to a filled version of the \ce{Ti2Ni}-type structure, with small nonmetallic atoms occupying the interstitial positions corresponding to the Wyckoff position $16d$ \cite{mackay1994new}. These atoms act as electron-acceptors, or anions. The structures of \ce{A4B2X} and \ce{A3B3X} differ by the occupation of the Wyckoff position $16c$. The chemically possible elemental site mutations of the $\eta$-carbide compounds offers a large tunability of their chemical and physical properties, while the effect and importance on the electronic properties of the light atoms in the void position has so far not been clarified.\cite{stover1956nickel,ettmayer1970thermische,conway2019interstitial}

$\eta$-Carbides have been found to be a versatile family of compounds for the investigation of emergent quantum properties \cite{ku1984new, waki2012magnetic,waki2011metamagnetism,waki2011interplay,ma2019superconductivity}, while only a few have been reported to be superconductors.\cite{ku1984new,ma2019superconductivity,matthias1963superconductivity} In a review by B.T. Matthias, \textit{et al.} inter alia superconductivity in \ce{Ti2Co}, \ce{Ti_{0.573}Rh_{0.287}O_{0.14}} and \ce{Ti_{0.573}Ir_{0.287}O_{0.14}} is mentioned, lacking detailed information.\cite{matthias1963superconductivity} The $\eta$-carbide superconductor with the highest known critical temperature of $T_{\rm c}$ = 9.7 K is \ce{Nb4Rh2C_{1-\delta}}.\cite{ku1984new,ma2021superconductivity} This material has recently been found to display a remarkably high upper critical field of $\mu_0 H_{\rm c2}$(0) = 28.5 T exceeding by far the weak-coupling Pauli limit.\cite{ma2021superconductivity} This material falls right in between the two commercially most often used superconducting materials. Namely, \ce{Nb4Rh2C_{1-\delta}} displays a critical temperature close to the one of NbTi, while having a much larger upper critical field, close to the one of \ce{Nb3Sn}.

Here, we report on the synthesis and superconducting properties of $\eta$-carbide type suboxides \ce{Ti4M2O} (M = Co, Rh, Ir). Single crystals of \ce{Ti4Co2O} and \ce{Ti2Co} were  prepared by an arc-melting and subsequent annealing process. By an elemental site substitution on the cobalt site in \ce{Ti4Co2O} with the other group-9  transition metals Rh and Ir, we obtained two new compounds \ce{Ti4Rh2O} and \ce{Ti4Ir2O}. Magnetic susceptibility, electrical resistivity  and  specific  heat  capacity  measurements show \ce{Ti4M2O} (M = Co, Rh, Ir) are bulk type-II superconductors with transition temperatures of 2.7, 2.8 and 5.4 K, respectively. All three compounds -- \ce{Ti4Co2O}, \ce{Ti4Rh2O} and \ce{Ti4Ir2O} -- display high upper critical field $H_{\rm c2}$(0) with respect to their critical temperatures. Our measurements show that \ce{Ti4Co2O} and \ce{Ti4Ir2O} even have upper critical fields above the weak-coupling BCS Pauli limit.  We, furthermore, show \ce{Ti2Co} is a semimetal with indications of a broad superconducting transition below 0.75 K. This indicates that incorporation of oxygen into the interstitial sites plays a crucial role for the electronic and superconducting properties of \ce{Ti4Co2O}. 

\section{Experimental Section}

\textbf{Single Crystals:} Single crystals of \ce{Ti2Co} and \ce{Ti4Co2O} were prepared from titanium powder (purity: 99.9 \%, Alfa Aesar), cobalt powder (purity: 99.99 \%, Strem Chemicals), and titanium dioxide powder (purity: 99.9 \%, Sigma-Aldrich). The reactants were thoroughly mixed in stoichiometric ratios and pressed into a pellet of a total mass of 600 mg. To avoid the oxidation of the metals, this process was performed in a glovebox under argon atmosphere. The pellet was melted in an arc furnace in a purified argon atmosphere on a water-cooled copper plate. The sample was molten 10 times in order to ensure optimal homogeneity. The solidified melt ball was sealed in a quartz ampule under 1/3 atm argon and annealed in a furnace for 30 days at 1000 $^\circ$C. Finally, the quartz tube was quenched in water. Large single crystals of \ce{Ti2Co} and \ce{Ti4Co2O} were obtained by mechanical breaking of the solidified ball (see photographs in supporting information).

\textbf{Polycrystalline Samples:} Polycrystalline samples of \ce{Ti4Rh2O} and \ce{Ti4Ir2O} were obtained from titanium powder (purity: 99.9 \%, Alfa Aesar), rhodium powder (99.99\%, Strem Chemicals), iridium powder (purity: 99.99 \%, Strem Chemicals), and titanium dioxide powder (purity: 99.9 \%, Sigma-Aldrich). Samples of a total mass of 120 mg were synthesized  from stoichiometric ratios of the reactants. We found the highest purity of \ce{Ti4Rh2O} was obtained with a  Ti:Rh:O ratio of 3.9:2.1:0.8, and \ce{Ti4Ir2O} was obtained with a Ti:Ir:O ratio of 4:2:0.7 in the arc-melting process. These were thoroughly mixed and pressed into a pellet. The pellet was melted in an arc furnace in a purified argon atmosphere on a water-cooled copper plate. The sample was molten 10 times in order to ensure optimal homogeneity. The solidified melt was ground to a fine powder and pressed into a pellet. The pellet was sealed in a quartz ampule under 1/3 atm argon and annealed in a furnace for 7 days at 1000 $^\circ$C. After the reaction, the quartz tube was quenched in water. Different starting stoichiometries in the arc-melting process have been tried to improve purity of the final polycrystalline samples.  

\textbf{Composition and Structure:}  The samples were investigated by means of powder X-ray diffraction (PXRD) measurements on a STOE STADIP diffractometer with Mo K${\alpha}$ radiation ($\lambda$  = 0.709300 Å). The PXRD patterns were collected in the 2$\theta$ range of 5-50$^{\circ}$  with a scan rate of 0.25$^{\circ}$/min. The PXRD of \ce{Ti4Rh2O} and \ce{Ti4Ir2O} samples were measured on a Rigaku diffractometer with Cu K${\alpha}$ radiation. Rietveld refinements were performed using the FULLPROF program suite \cite{rodriguez1993recent}. Single crystal X-ray diffraction (SXRD) was performed on a Rigaku's XtaLAB Synergy diffractometer using Cu or Mo K${\alpha}$ radiation. The graphical interface Olex2 \cite{dolomanov2009olex2} was used to solve the structure and do the refinement. The structure was solved using intrinsic phasing provided by the ShelXT program \cite{sheldrick2015shelxt} and refined by the least squares method of ShelXL \cite{sheldrick2015crystal}.

\textbf{Physical Properties:} The temperature-dependent magnetization measurements were performed using a Quantum Design Magnetic Properties Measurement System (MPMSXL) equipped with a reciprocating sample option (RSO). The measured plate like samples were placed in parallel to the external magnetic field to minimize demagnetization effects. Specific heat capacity and resistivity measurements were performed with a Quantum Design Physical Property Measurement System (PPMS) using a He-3 insert option for measurements below 2 K. Temperature-dependent resistivity data of \ce{Ti2Co} and \ce{Ti4Co2O} were measured on single crystals by employing the standard 4-probe technique. Temperature-dependent resistivities of \ce{Ti4Rh2O} and \ce{Ti4Ir2O} were measured on sintered polycrystalline pellets using the same technique. For all resistivity measurements, silver wires were connected to the sample with silver paint.

\textbf{Electronic Structure Calculations:} We have performed density functional theory calculations for all compounds using the full potential local orbital (FPLO) basis~\cite{Koepernik1999} and the generalized gradient approximation (GGA) to the exchange correlation functional~\cite{Perdew1996}. Due to the importance of spin-orbit coupling for Co, Rh and Ir, we have performed fully relativistic calculations. The experimentally determined crystal structures were used for all calculations. 

\section{Results and discussion}

\subsection{Structure of \ce{Ti4M2O} (with M = Co, Rh, Ir)}

Large single crystals of \ce{Ti4Co2O} with shiny metallic luster were prepared by arc-melting and subsequent annealing (see supporting information). The structure of \ce{Ti4Co2O} was previously reported in the literature, but only low quality polycrystalline data was available so far \cite{ku1984new}. The SXRD data obtained here allows us to provide a high-quality structural solution in table \ref{table:Table1}. 

\begin{figure}
\centering
\includegraphics[width= 0.5\textwidth]{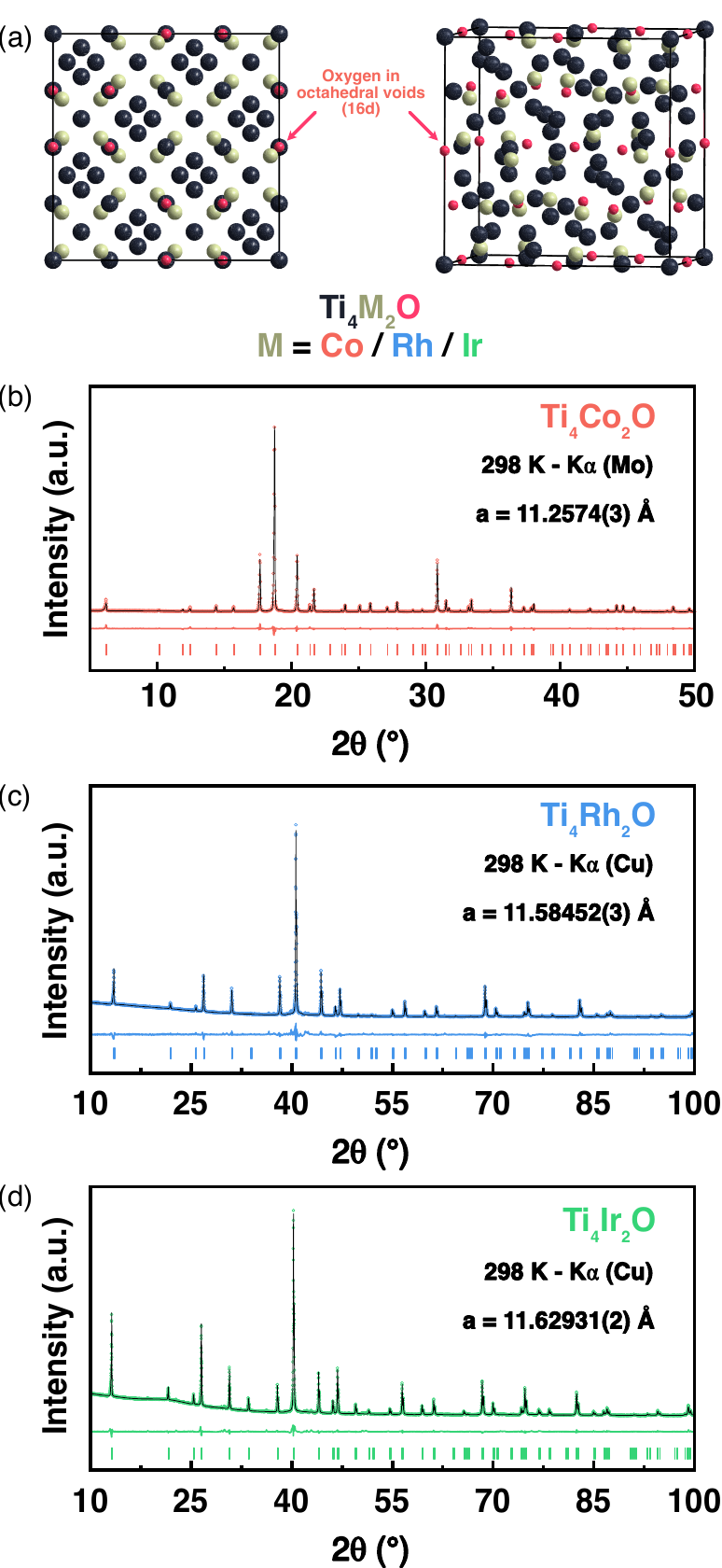}
\caption{(a) Schematic representation of two orientations for the refined crystal structure from SXRD of \ce{Ti4Co2O}. PXRD patterns and Rietveld refinements of (b) \ce{Ti4Co2O}, (c) \ce{Ti4Rh2O} and (d) \ce{Ti4Ir2O}, respectively.  The plots are represented as follows: observed (colored dots), calculated (black line) and difference (colored line) intensities. The Bragg positions of each phase are indicated with colored vertical bars.}
\label{fig:figure1}
\end{figure}

\ce{Ti4Co2O} crystallizes in the cubic $Fd\bar{3}m$ space group, with the cell parameter $a =  11.2574(3)$ $\si{\angstrom}$. In figure \ref{fig:figure1}(a), we show a schematic view of the refined crystal structure of \ce{Ti4Co2O}. The composition for \ce{Ti4Co2O} was found to be fully stoichiometric, with no oxygen vacancies. Temporary freed site occupancy showed no disorder over the different sites, and the low electronic residues confirm that all positions are correctly filled with the right atoms. All the details of the SXRD and the structural refinements are summarized in table \ref{table:Table1}. The whole sample was found to be phase pure as evidenced by the Rietveld refinement shown in figure \ref{fig:figure1}(b).

\begin{table}
\centering
\caption{Crystallographic data for single-crystals of \ce{Ti2Co} and \ce{Ti4Co2O}.}
\label{table:Table1}
\begin{tabular}{c c c} 
\hline 
 
 Formula & \ce{Ti2Co} & \ce{Ti4Co2O} \\ 
 Formula weight ($g \, mol^{-1}$) & 154.73 & 325.46 \\
 Crystal system & \multicolumn{2}{c}{ Cubic} \\
 Space group & \multicolumn{2}{c}{Fd$\bar{3}$m (227)}\\
 a, b, c (Å) & 11.2865(5) & 11.2574(3)\\
 $\alpha$, $\beta$, $\gamma$ ($^{\circ}$) & \multicolumn{2}{c}{90} \\
Volume (Å$^3$) & 1437.73(19) & 1426.63(11) \\
Z & 32 & 16 \\
Calculated density ($g \, cm^{-3}$)& 5.719 & 6.061\\
Temperature ($K$) & \multicolumn{2}{c}{160} \\
 Radiation & {Cu K$\alpha$ (1.54184 $\si{\angstrom}$)} & {Mo K$\alpha$ (0.71073 $\si{\angstrom}$)}\\
 Crystal color & \multicolumn{2}{c}{Metallic silver-black} \\
 Crystal size ($mm^3$) & 0.064 $\times$ 0.041 $\times$ 0.037 & 0.12 $\times$ 0.11 $\times$ 0.07 \\
 Linear absorption coefficient ($cm^{-1}$) & 1406.66 (Cu) & 173.75 (Mo)\\
 Scan mode & \multicolumn{2}{c}{$\omega$} \\
 2$\theta$ range for data collection ($^{\circ}$) & 13.63 --- 137.992 & 7.206 --- 64.518\\
 \multirow[t]{3}{*}{Index ranges} & -5 $\leq$ h $\leq$ 13 & -16 $\leq$ h $\leq$ 13 \\
  & -12 $\leq$ k $\leq$ 8 & -15 $\leq$ k $\leq$ 16\\
  & -15 $\leq$ l $\leq$ 16 & -12 $\leq$ l $\leq$ 9 \\  
   Reflections collected & 278 & 1100 \\ 
 
 \multicolumn{3}{c}{\textbf{Data reduction}} \\
 Completeness (\%) & 100 & 99.31 \\
  Numbers of independent reflections & 84 & 129 \\
  $R_{\rm int}$ (\%) & 2.88 & 2.05 \\
Empirical absorption correction & \multicolumn{2}{c}{Auto frame scaling, sample decay, 4x4 detector correction} \\
Numerical absorption correction & Spherical & Analytical \\
 Independent reflections with I $\geq$ 2.0$\sigma$ & 76 & 128 \\ [4ex]

  \multicolumn{3}{c}{\textbf{Refinement}}\\
 R$_{1}$ (obs / all) (\%) & 3.07 / 3.24 & 1.23 / 1.23   \\
 wR$_{2}$ (obs / all) (\%) & 8.50 / 8.57 & 3.15 / 3.15 \\
 GOF & 1.130 & 1.192 \\
 Numbers of refined parameters & 10 & 13 \\
 Difference Fourier residues ($e^- \, \si{\angstrom}^{-3}$) & -0.436 --- +0.505 & -0.519 --- +0.319  \\
  \hline
\end{tabular}
\end{table}

The structure of \ce{Ti4Co2O} corresponds to a filled version of the \ce{Ti2Ni} structure with the oxygen atoms occupying the $16d$ Wyckoff positions. In the SXRD refinement of \ce{Ti4Co2O}, an electronic density residue of 19.48 $e^- \, \si{\angstrom}^{-3}$ is found at the cell center, if the oxygen is not considered for the crystallographic void positions, evidencing the incorporation of oxygen in the structure. Here, titanium atoms occupy the $16c$ and the $48f$ Wyckoff positions, cobalt atoms occupy the $32e$ Wyckoff positions, while oxygen atoms occupy the $16d$ Wyckoff positions, resulting in a formula of \ce{Ti64Co32O16} for one unit cell. The \ce{Ti2Ni}-type structures with space group of $Fd\bar{3}m$ contain in their unit cell 96 metal atoms.  Alternatively, the large \ce{Ti2Ni} unit cell can also be understood as eight cubic sub-cells having two alternating patterns, and the interstitial positions between these sub-cells are empty. \cite{ma2021superconductivity}. The \ce{Ti2Ni} structure-type is well known to act as a host with some tolerance, allowing the incorporation of light elements, such as oxygen, nitrogen and carbon, in the interstitial positions to form the $\eta$-carbide structure. 

Polycrystalline samples of the two suboxides \ce{Ti4Rh2O} and \ce{Ti4Ir2O} were obtained by means of solid-state synthesis. The PXRD patterns and their respective Rietveld refinements are shown in figure \ref{fig:figure1}(c)\&(d). \ce{Ti4Rh2O} and \ce{Ti4Ir2O} crystallize in the same structure as \ce{Ti4Co2O}, hence in the cubic space group $Fd\bar{3}m$ with cell parameters $a$ = 11.58452(3) \AA \ and $a$ = 11.62931(2) \AA, respectively. The unit cell increases with increasing atomic radii within this series of compounds, as it is expected. The presented Rietveld refinements lead to reliability factors $R_{wp}$ of 4.10 \% and 4.19 \% for \ce{Ti4Rh2O} and \ce{Ti4Ir2O}, respectively. Hence, the refined structures of the two new compounds, \ce{Ti4Rh2O} and \ce{Ti4Ir2O} are of excellent quality. Details on the Rietveld refinements are summarized in table \ref{table:Table2}. In table \ref{table:Table3} the atomic coordinates and the atomic displacement parameters for all three compounds are assembled. All measurements were performed on these polycrystalline samples. However, the validity of the structural model could also be confirmed by SXRD measurements, the details of which are presented in the Supporting Information. The cell parameters of the SXRD and PXRD measurements are in excellent agreement.

\begin{table}
\centering
\caption{Details on the Rietveld refinement for polycrystalline samples of \ce{Ti4Rh2O} and \ce{Ti4Ir2O}.}

 \begin{tabular}{c c c} 
 \hline 
 
 Formula & \ce{Ti4Rh2O} & \ce{Ti4Ir2O} \\ 
 Formula weight ($g \, mol^{-1}$) & 413.33 & 591.96 \\
 Crystal system & \multicolumn{2}{c}{ Cubic} \\
 Space group & \multicolumn{2}{c}{Fd$\bar{3}$m (227)}\\
 a, b, c (Å) & 11.58452(3) & 11.62931(2)\\
 $\alpha$, $\beta$, $\gamma$ ($^{\circ}$) & \multicolumn{2}{c}{90} \\
Volume (Å$^3$) & 1554.656(7) & 1572.759(6) \\
Z & \multicolumn{2}{c}{16} \\
Calculated density ($g \, cm^{-3}$)& 7.06359 & 9.999\\
Diffractometer & \multicolumn{2}{c}{Rigaku SmartLab, Bragg-Brentano}\\
Temperature ($K$) & \multicolumn{2}{c}{298} \\
  Radiation & \multicolumn{2}{c}{Cu K$\alpha$ (1.54184 $\si{\angstrom}$)}\\
 2$\theta$ range (deg.), no. of points & 5.07 - 100.01, 9495  & 5.08 - 100.02, 9495 \\
 Step size (deg), counting time
(s) & \multicolumn{2}{c}{0.01, 5.99766} \\
 Background & \multicolumn{2}{c}{Linear interpolation}\\
 Profile function & \multicolumn{2}{c}{\makecell{Thompson-Cox-Hastings pseudo-Voigt\\with Axial divergence asymmetry}}\\
 No. of refined parameters & \multicolumn{2}{c}{11} \\
 $R_B$ (\%)& 6.22 & 3.88 \\
 $R_p$ (\%)& 2.49 & 2.41 \\
 $R_{wp}$ (\%)& 4.10 & 4.19 \\
 $R_{exp}$ (\%) & 1.38 & 1.09 \\
GOOF & 2.9649 & 3.8487\\
   \hline
\end{tabular}
\label{table:Table2}
\end{table}

\begin{table}
\centering
\caption{Refined coordinates, atomic displacement parameters (ADPSs) (\AA$^2$), and their estimated standard deviations for \ce{Ti2Co}, \ce{Ti4Co2O}, \ce{Ti4Rh2O}, and \ce{Ti4Ir2O}.}
\label{table:Table3}
 \begin{tabular}{c c c c c c c} 
 \hline 
 
 & \makecell{Wyckoff\\position} & & \ce{Ti2Co} & \ce{Ti4Co2O} & \ce{Ti4Rh2O} & \ce{Ti4Ir2O} \\ 
 
 x(Ti1) & 16c & & \multicolumn{4}{c}{0}\\
 & & $U_{iso}$ & 0.0517(8) &	0.0049(2) &	0.0296(8) & 0.0163(12)\\
 
 x(Ti2) & 48f & & 0.43659(14)&0.43890(5)&0.44217(8)&0.44086(13)\\
 & & $U_{iso}$ & 0.0437(7)&0.00385(18)&0.0261(3)&0.0197(5)\\
 
 x(M) & 32e & & 0.21406(10)&0.21131(2)&0.21492(3)&0.21468(3)\\
& & $U_{iso}$ &0.0537(8)&0.00496(18)&0.0292(2)&0.01342(19)\\

 x(O) & 16d & & - & \multicolumn{3}{c}{1/2}\\
 & & $U_{iso}$ &- & 0.0045(7)&0.016(2)&0.057(5)\\

   \hline
\end{tabular}
\end{table}

\subsection{Superconductivity in \ce{Ti4M2O} (with M = Co, Rh, Ir)}

\begin{figure}
 \centering
\includegraphics[height = 0.85\textheight]{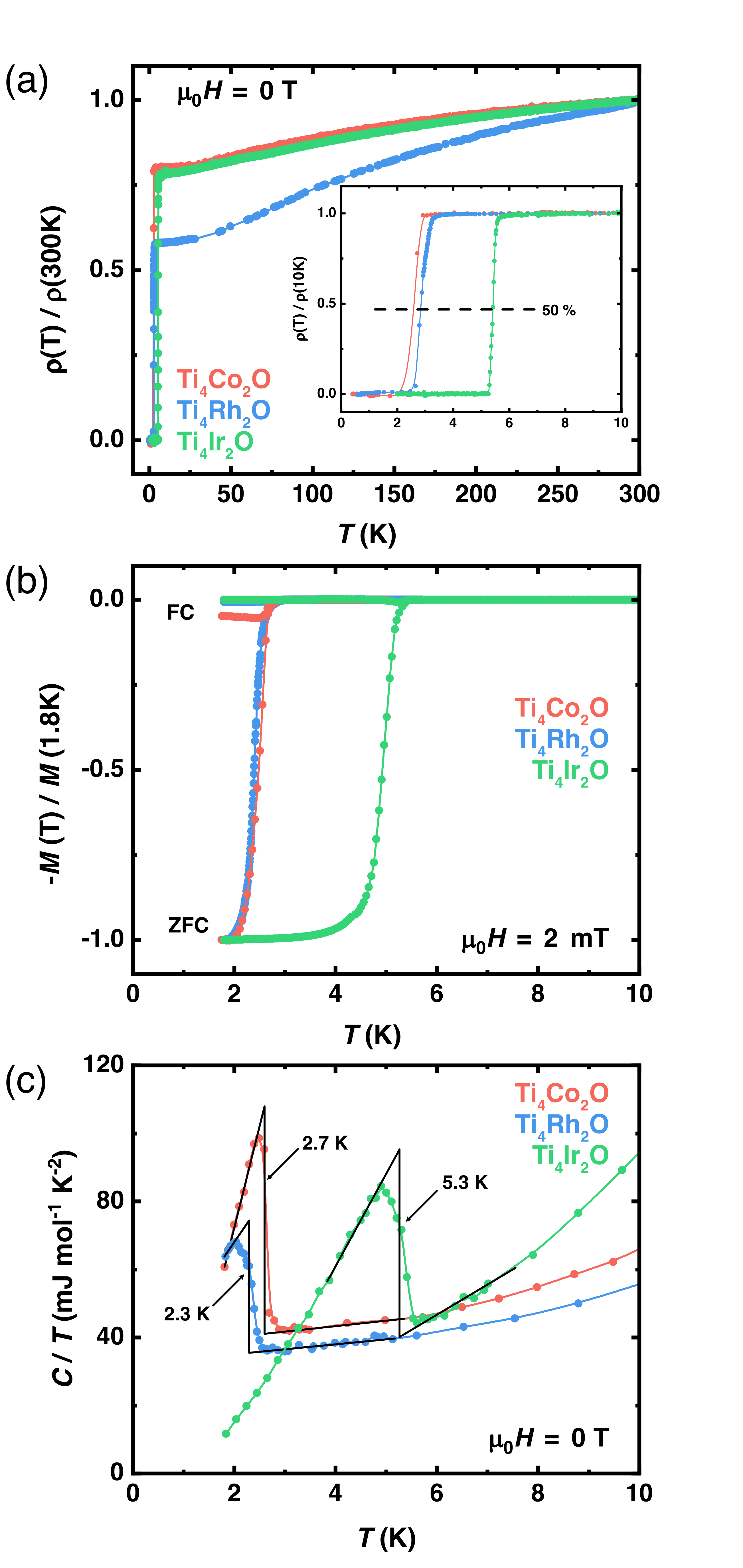}
\caption{Superconducting properties of \ce{Ti4M2O} with M = Co, Rh, Ir: (a) Temperature-dependent resistivities in zero-field in the temperature range of $T =$ 300 K and 400 mK, the inset show the resistivity in the vicinity of the superconducting transition (b) ZFC and FC magnetization in an external field of $\mu_0 H$ = 2 mT (c) Temperature-dependent specific heat capacity as $C$/$T$ with entropy conserving constructions in zero-field.}
\label{fig:figure2}
\end{figure}

We have performed temperature-dependent electrical resistivity measurements on single crystals of \ce{Ti4Co2O}, and on polycrystalline samples of \ce{Ti4Rh2O} and \ce{Ti4Ir2O}. In figure \ref{fig:figure2}(a), we show resistivity measurements in the temperature range between $T =$ 300 K and 400 mK. All three compounds are found to be metals with residual resistivity ratios (RRR) -- here defined as $\rho$(300 K)/$\rho$(10 K) -- of RRR = 1.26, 1.72, and 1.27 for \ce{Ti4Co2O}, \ce{Ti4Rh2O}, and \ce{Ti4Ir2O}, respectively. 

These values correspond to a poor metal behavior, as commonly observed in metal oxides (see, e.g., reference \citenum{von2019h}). The room-temperature resistivities of \ce{Ti4Co2O}, \ce{Ti4Rh2O}, and \ce{Ti4Ir2O} are very similar, with ${\rho(300K)}$ = 0.07 m$\Omega$ cm, ${\rho(300K)}$ = 0.12 m$\Omega$ cm, and ${\rho(300K)}$ = 0.25 m$\Omega$ cm, respectively. 

We find that all three compounds clearly undergo a transition to a superconducting state at critical temperatures of $T_{\rm c}$ = 2.7 K, 2.8 K, and 5.4 K for \ce{Ti4Co2O}, \ce{Ti4Rh2O}, and \ce{Ti4Ir2O}, respectively. These transitions are manifested by the sharp drop in the resistivity at the corresponding temperatures. In the inset of \ref{fig:figure2}(a), the electrical resistivities of the three compounds in the vicinity of the superconducting transition are depicted. The observation of superconductivity in \ce{Ti4Co2O} is in agreement with earlier findings, where a critical temperature $T_{\rm c}$ of 3.10-2.84 K was enlisted for this compound, in the absence of any further experimental details (see, reference \citenum{ku1984new}).

The superconducting state of these materials was further characterized with magnetization and heat capacity measurements. In figure \ref{fig:figure2}(b), we show the temperature-dependent magnetic susceptibilities (-$M$/$M$(1.8 K)) of \ce{Ti4Co2O}, \ce{Ti4Rh2O}, and \ce{Ti4Ir2O} in zero-field cooled (ZFC) and field-cooled (FC) conditions in an external magnetic field of $\mu_0 H$ = 2 mT. The observed superconducting transition temperatures in the magnetization are in good agreement with the critical temperatures from the resistivity measurements. We observe shielding fractions clearly larger than $\chi =$ -1 for all three compounds, caused by demagnetization effects. In the normal state, all three compounds are found to be Pauli paramagnets (see Supporting Information).

In figure \ref{fig:figure2}(c), we present the temperature-dependent specific heat $C$/$T$ vs $T$  for \ce{Ti4Co2O}, \ce{Ti4Rh2O} and \ce{Ti4Ir2O}. In the normal state, the specific heat consists of an electronic ($\rm C_{el}$) and a phononic ($\rm C_{ph}$) contribution, it was fitted according to the following relationship:
 
\begin{equation}
\frac{C(T)}{T} = \frac{C_{el} + C_{ph}}{T} = \gamma + \beta T^2.
\label{eq:specific}
\end{equation}

Here $\gamma$ is the Sommerfeld coefficient and $\beta$ is the coefficient of the phonon contribution. The value for $\gamma$ is found to be $\gamma =$ 39.98 mJ mol$^{-1}$ K$^{-2}$  for \ce{Ti4Co2O}, $\gamma$ = 34.5 mJ mol$^{-1}$ K$^{-2}$ for \ce{Ti4Rh2O}, and $\gamma$ = 29.3 mJ mol$^{-1}$ K$^{-2}$ for \ce{Ti4Ir2O}. $\gamma$ is proportional to the electronic density of states at the Fermi-level $D$($E_{\rm F}$). Hence, it can be argued that all three compounds have very similar $D$($E_{\rm F}$).

The phononic coefficients are $\beta$ =  0.24 mJ mol$^{-1}$ K$^{-4}$ for \ce{Ti4Co2O}, $\beta$ = 0.22 mJ mol$^{-1}$ K$^{-4}$ for \ce{Ti4Rh2O}, and $\beta$ = 0.59 mJ mol$^{-1}$ K$^{-4}$ for \ce{Ti4Ir2O}. The Debye temperatures $\Theta_{\rm D}$ can be determined according to the equation
 
\begin{equation}
\Theta_D = \left(\frac{12 \pi^4}{5 \beta} n R \right)^{\frac{1}{3}}
\label{eq:Theta}
\end{equation}
where n is the number of atoms per formula unit, and R = 8.314 J mol$^{-1}$ K$^{-1}$ is the ideal gas constant. The Debye temperature is calculated to be $\Theta_{\rm D} =$ 386 K for \ce{Ti4Co2O}, $\Theta_{\rm D} =$ 397 K for \ce{Ti4Rh2O}, and $\Theta_{\rm D} =$ 285 K for \ce{Ti4Ir2O}. 

The clearly distinct discontinuities in the specific heat are in good agreement with the superconducting transition temperatures observed in the magnetization and resistivity measurements. With the entropy-conserving construction shown in figure \ref{fig:figure2}(c), the normalized specific heat jump $\Delta C/\gamma T_{\rm c}$ was determined. $\Delta C/\gamma T_{\rm c}$ is found to be 1.65 for \ce{Ti4Co2O}, 1.28 for \ce{Ti4Rh2O} and 1.80 for \ce{Ti4Ir2O}. All three values are close to the BCS weak-coupling value for $\Delta C/\gamma T_{\rm c}$ of 1.43, confirming the bulk nature of the superconductivity in these materials. Therefore, all three compounds are bulk suboxide-superconductors. 

For \ce{Ti4Ir2O}, we have furthermore determined the lower critical field $\mu_0 H_{\rm c1}$(0) by employing a series of ZFC field-dependent magnetization measurements in the superconducting state (see Supporting Information). Here, we take the magnetic field point where the $M$($H$) curve first deviates from linearity as the measure for $\mu_0 H_{\rm c1}$(T) \cite{naito1990temperature}. With this approximation, the obtained $\mu_0 H_{\rm c1}$(T) values are fitted using the empirical equation $H_{\rm c1}(T) = H_{\rm c1}(0) [1-(T/T_{\rm c})^2]$. We determine the lower critical field for \ce{Ti4Ir2O} to be $\mu_0 H_{\rm c1}$(0) = 8.9 mT (see Supporting Information). 

\subsection{High-upper critical fields in \ce{Ti4M2O} (with M = Co, Rh, Ir)}

\begin{figure}
 \centering
\includegraphics[width= \textwidth]{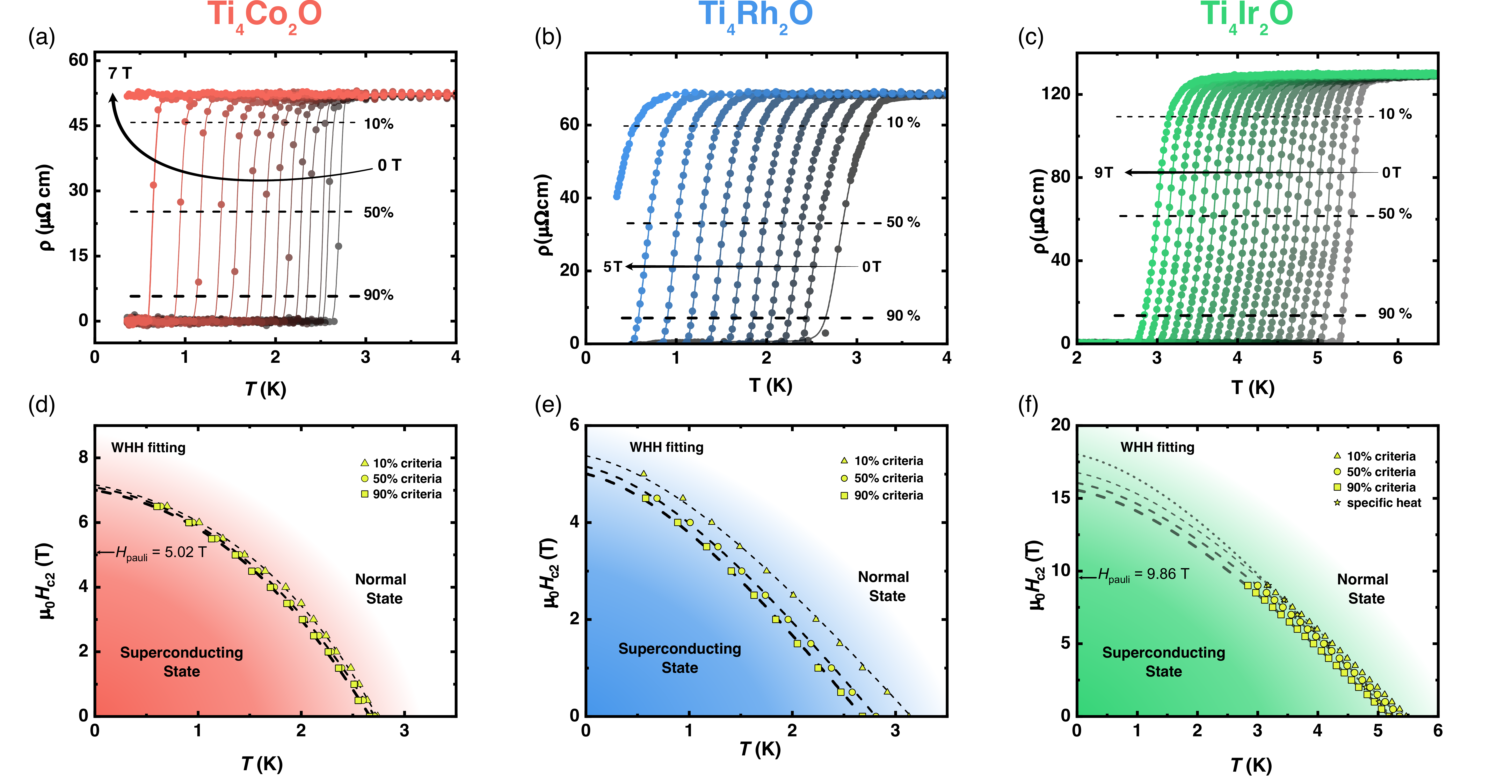}
\caption{(a)-(c) Field-dependent and temperature-dependent resistivity of \ce{Ti4M2O} with M = Co, Rh and Ir. (d)-(f) Electronic phase diagrams of \ce{Ti4M2O} with M = Co, Rh and Ir.}
\label{fig:figure3}
\end{figure}

In figure \ref{fig:figure3}(a)-(c) we show temperature- and field-dependent resistivity measurements in the vicinity of the superconducting transitions on a single crystal of \ce{Ti4Co2O} and on polycrystalline samples of \ce{Ti4Rh2O} and \ce{Ti4Ir2O}. As expected, the critical temperature decreases steadily as the applied magnetic field increases. However, the shift of the superconducting transition temperatures in these materials for all three compounds in external fields is remarkably small. Commonly, conventional superconductors have upper critical fields clearly below the weak-coupling Pauli limit. Hence, should be suppressed at the here applied fields. 

For \ce{Ti4Co2O} shown in figure \ref{fig:figure3}(a), the critical temperature in the resistivity is still $T_{\rm c}$ = 0.65 K in an external field of $\mu_0 H$ = 6.5 T. The weak-coupling Pauli limit of \ce{Ti4Co2O} is determined to be $\mu_0 H_{Pauli}$ = 5.0 T. In \ce{Ti4Ir2O} the critical temperature in a $\mu_0 H$ = 9 T field is still $T_{\rm c}$ = 3.0 K. The superconductivity is normally long suppressed in such high magnetic fields in comparable materials \cite{gornicka2019iridium, xiao2021normal}. Also, the remarkable sharpness of the transition in high fields is worth emphasizing. The transition width in the resistivity of \ce{Ti4Ir2O} in zero-field is $\Delta T_{\rm c}$ = 0.34 K, and in a magnetic field of  $\mu_0 H$ = 9 T it is still $\Delta T_{\rm c}$ = 0.76 K. Such a small widening of the transition is quite remarkable and indicates a wide vortex solid region, and a small vortex liquid region. A feature especially of importance for the realization of superconducting magnet applications \cite{sharma2021superconductivity}.

Here, we take the common 10\%-, 50\%- and 90\% criteria of resistances to determine the upper critical field at zero temperature $\mu_0H_{\rm c2}$(0). In figure \ref{fig:figure3}(d)-(f), we estimate the upper critical fields at zero temperature $\mu_0H_{\rm c2}$(0) using the Werthamer-Helfand-Hohenberg (WHH) approximation in the clean limit \cite{baumgartner2013effects}:
    
\begin{equation}
\mu_0 H_{c2} (T)= \frac{\mu_0 H_{c2} (0)}{0.73} h^\ast_{fit} (T/T_{\rm c}).
\label{eq:WHH}
\end{equation}
\begin{equation}
 h^\ast_{fit} (t)=(1-t)-C_{1}(1-t)^2-C_{2}(1-t)^4.
\end{equation}

This approach can be considered as a conservative estimation of the upper critical fields $\mu_0H_{\rm c2}$(0). The fitting gives an upper critical value of $\mu_0 H_{\rm c2}$(0)= 7.18 T, 7.08 T, and 7.02 T for the 10\%-, 50\%- and 90\% criteria, for \ce{Ti4Co2O} respectively. All of these values are clearly higher than weak-coupling Pauli limit. For the estimated upper critical field values of \ce{Ti4Rh2O} we find $\mu_0H_{\rm c2}$(0) = 5.38 T, 5.15 T, and 5.01 T for the 10\%-, 50\%- and 90\% criteria, respectively. which is large, and very close to the weak-coupling Pauli limit, where $\mu_0 H_{\rm Pauli}$ = 5.21 T for the 50\%-criteria. 

The estimated upper critical field values from the 10\%-, 50\%-, 90\%-criteria and specific heat capacity for \ce{Ti4Ir2O} are $\mu_0H_{\rm c2}$(0) = 16.76 T, 16.06 T, 15.47 T, and 18.03 T which is remarkably high. This upper critical field is by far exceeding the weak-coupling Pauli limit of $\mu_0 H_{\rm Pauli}$ = 9.86 T for the 50\%-criteria. 

The observation of the violation of the weak-coupling Pauli limit in these materials is remarkable. These cubic, centrosymmetric materials -- without any local magnetic moments -- would from a material design perspective not to be expected to display such uncommon superconducting properties. However, \ce{Ti4Co2O} and \ce{Ti4Ir2O} are also two additional examples for the observation of the weak-coupling Pauli limit violation after the recent report on \ce{Nb4Rh2C_{1-\delta}} in the same structure type.\cite{ma2021superconductivity} These similar properties occur in this structure type, although there are significant differences in chemical composition. Not only the transition elements are different, but also the anionic void filling element is changed to oxygen. 

\subsection{Parameters of the superconductivity in \ce{Ti4M2O} (M = Co, Rh, Ir)}

According to the Ginzburg-Landau theory, the obtained $\mu_0 H_{c2}$(0) value (here the values obtained from the 50\% criteria are being used) can be used to estimate the coherence length $\xi_{\rm GL}$(0) at T = 0 K using the following equation:

\begin{equation}
\mu_0 H_{c2}(0) = \frac{\Phi_0}{2 \pi \ \xi_{\rm GL}^2}.
\label{eq:GL}
\end{equation}

Here, $\Phi_0 = h/(2e) \approx 2.0678 \cdot 10^{-15}$ Wb is the quantum flux. The calculated coherence length $\xi_{\rm GL}$(0) is 68.2 \AA \  for \ce{Ti4Co2O}, 79.9 \AA \  for \ce{Ti4Rh2O}, and 45.3 \AA \  for \ce{Ti4Ir2O}, respectively. The superconducting penetration depth $\lambda_{\rm GL}$ of \ce{Ti4Ir2O} was estimated from the obtained values of $\xi_{\rm GL}$ and $H_{c1}$ by using the relation:
 
\begin{equation}
\mu_0 H_{c1} = \frac{\Phi_0}{4 \pi \lambda_{\rm GL}^2} ln(\frac{\lambda_{\rm GL}}{\xi_{\rm GL}}).
\end{equation}
We obtained a value of $\lambda_{\rm GL} =$ 2756 \AA \  for \ce{Ti4Ir2O}. The value of $\kappa_{\rm GL} = \lambda_{\rm GL}/\xi_{\rm GL} =$ 60.8. 

The electron-phonon coupling constant $\lambda_{\rm ep}$ of \ce{Ti4M2O} can be  estimated from the Debye temperature, using the semi-empirical McMillan approximation \cite{mcmillan1968transition}:
  
\begin{equation}
\lambda_{\rm ep} = \dfrac{1.04 + \mu^{*} \ {\rm ln}\big(\frac{\Theta_{\rm D}}{1.45 T_{\rm c}}\big)}{(1-0.62 \mu^{*}) {\rm ln}\big(\frac{\Theta_{\rm D}}{1.45 T_{\rm c}}\big)-1.04}.
\label{eq:McM}
\end{equation} 

The Coulombic repulsion parameter $\mu^{*}$ is set to be 0.13 according to the empirical approximation for similar materials \cite{stolze2018sc,von2014superconductivity,von2016effect}. Based on the obtained value, the electron-phonon coupling constant is calculated to be $\lambda_{\rm ep}$ = 0.51 for \ce{Ti4Co2O}, $\lambda_{\rm ep}$ = 0.52 for \ce{Ti4Rh2O}, and $\lambda_{\rm ep}$ = 0.66 for \ce{Ti4Ir2O}. Using these electron-phonon coupling constants and the Sommerfeld constants, we can estimate the density of states at the Fermi-level by using the following relationship: 

\begin{equation}
D(E_{\rm F}) = \dfrac{3 \gamma}{\pi^2 k_{\rm B}^2 (1+\lambda_{\rm ep})}.
\label{eq:DOS}
\end{equation}

This results in a density of states at the Fermi-level of $D(E_{\rm F})$ = 11.22 states eV$^{-1}$ per formula unit (f.u.) for \ce{Ti4Co2O}, $D(E_{\rm F})$ = 9.64 states eV$^{-1}$ per formula unit (f.u.) for \ce{Ti4Rh2O}, and $D(E_{\rm F})$ = 7.50 states eV$^{-1}$ per formula unit (f.u.) for \ce{Ti4Ir2O}.

It can be stated that all three superconductors \ce{Ti4M2O} (M=Co, Rh, Ir) have very high upper critical fields compared with their $T_{\rm c}$ values. We find that the upper critical fields of \ce{Ti4Co2O} and \ce{Ti4Ir2O} are even exceeding the commonly expected maximal upper critical field, known as the weak-coupling Pauli limit. By analyzing the electronic and phononic contribution, there is no obvious experimental observation that might allow to explain this unusual electronic effect. That both \ce{Ti4Co2O} and \ce{Ti4Ir2O} are violating this limit is a strong indicator that the most commonly expected cause for this effect -- strong spin-orbit coupling -- is not the main reason for the observation at hand. Since the effect is observed for the lighter cobalt, as well as for the heavier iridium compound, even though the spin orbit-coupling is enhanced substantially from one to the other element. A comprehensive comparison of the electronic parameters of the three $\eta$-carbide-type suboxide superconductors is listed in table \ref{table:Table4}.

\begin{table}
\centering
\caption{Summary of all the determined parameters of \ce{Ti4M2O} (M=Co, Rh, Ir).}

 \begin{tabular}{c c c c c} 
 \hline 
 
 Parameter & Units & \ce{Ti4Co2O} & \ce{Ti4Rh2O} & \ce{Ti4Ir2O} \\ 
 
 \hline
 $T_{\rm c,magnetization}$ & K &  2.7 &  2.7 & 5.3 \\ 
 $T_{\rm c,resistivity}$ & K &  2.7 &  2.8 & 5.4 \\ 
$T_{\rm c,specific heat}$ & K & 2.7 &  2.3 & 5.3 \\ 
$\rho$(300K) &  m$\Omega$ cm & 0.07 &  0.12 & 0.25  \\ 
RRR  & - &  1.26 &  1.72 & 1.27  \\ 
$\mu_0 H_{\rm c1}(0)$ &  mT  &  - & - & 8.9  \\ 
$\mu_0 H_{\rm c2}(0)$  &  T &  7.08 &  5.15  & 16.06 \\ 
$\beta$  &  mJ mol$^{-1}$ K$^{-4}$ &  0.236 &  0.217 & 0.587  \\ 
$\gamma$  &  mJ mol$^{-1}$ K$^{-2}$ &  39.9 &  34.5  & 29.3   \\ 
$\Theta_{\rm D}$  &  K &  386 &  397 & 285\\ 
$\xi_{GL}$  &  \AA &  68.2 &  79.9 & 45.3 \\
$\lambda_{GL}$ &  \AA &  - & -  & 2756   \\
$\kappa_{\rm GL}$ &  - &  - & - & 60.8   \\
$\Delta C/\gamma T_{\rm c}$  & - &  1.65 &  1.28 & 1.80  \\ 
$\lambda_{\rm ep}$ &-& 0.51 &  0.52 & 0.66 \\ 
$D$($E_{\rm F}$) &  states eV$^{-1}$ per f.u.&   \ 11.22  &  9.64 & 7.50 \\ 
$\mu_0H_{\rm Pauli}$ &  T &  5.02  &  5.21  & 9.86  \\ 
$\mu_0H_{\rm c2}(0)/T_{c}$  &   T/K  &  2.62 &  1.84 & 3.03 \\

   \hline
\end{tabular}
\label{table:Table4}
\end{table}

\subsection{The structure and electronic properties of \ce{Ti2Co}}

\begin{figure}
\centering
\includegraphics[width= \textwidth]{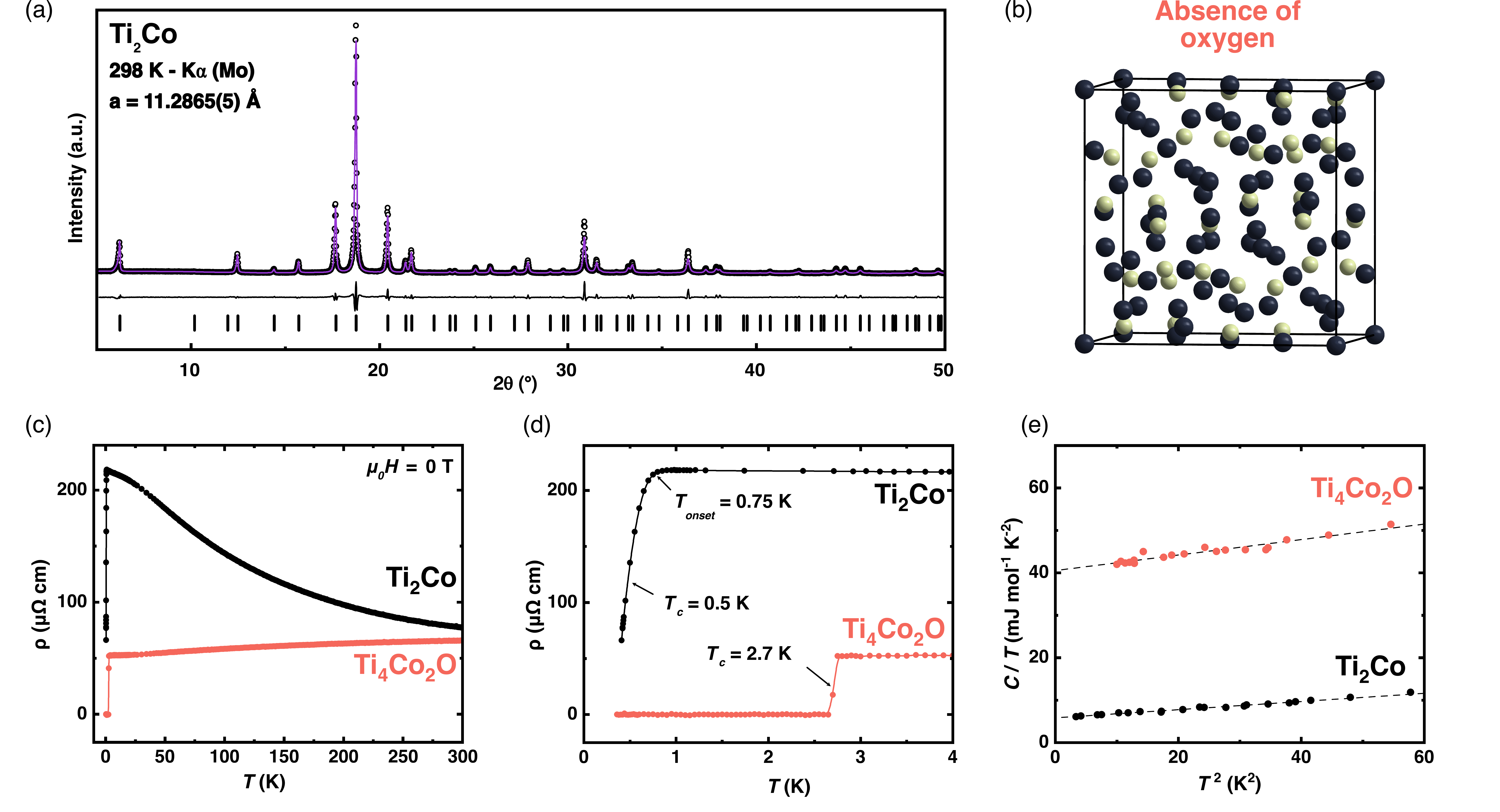}
\caption{Structure and electronic properties of \ce{Ti2Co}: (a) PXRD pattern and Rietveld refinement. (b) Schematic representation of the \ce{Ti2Co} structure as obtained from SXRD refinement. (c) Temperature-dependent resistivities of \ce{Ti2Co} and \ce{Ti4Co2O} in zero-field, and (d) zoom-in on the low-temperature region of the resistivities. (e) Normal-state specific heat of \ce{Ti2Co} and \ce{Ti4Co2O}.}
\label{fig:figure4}
\end{figure}

Large single crystals of \ce{Ti2Co} with a silver metallic luster were obtained analogously to the synthesis of \ce{Ti4Co2O} single crystals (see supporting information). The structure of \ce{Ti2Co} had previously been reported\cite{ku1984new}, here we solve it for the first time using high-quality SXRD data. The PXRD pattern of the finely ground crystals is shown in figure \ref{fig:figure4}(a), evidencing the absence of any impurity phases. In figure \ref{fig:figure4}(b), we show a schematic view of the refined crystal structure of \ce{Ti2Co}.

\ce{Ti2Co} crystallizes in the \ce{Ti2Ni} structure in the cubic $Fd\bar{3}m$ space group, with the cell parameter $a =  11.2865(5)$ $\si{\angstrom}$. This unit cell is slightly larger than the above-mentioned cell parameter for \ce{Ti4Co2O}, this is in good agreement with the increased ionicity of the bonding upon oxygen incorporation. In our \ce{Ti2Co} crystals, there are no atoms in the void position, as evidence by the absence of any  significant electronic residue in the SXRD refinement of \ce{Ti2Co}. All the details of the SXRD and the structural refinements are summarized in table \ref{table:Table1}. Attempts to obtain isostructural oxygen-free \ce{Ti2Rh} and \ce{Ti2Ir} were unsuccessful. \ce{Ti2Rh} crystallizes in the \ce{Zr2Cu}-type structure (space group: $I4/mmm$ (139)) \cite{wodniecki2007time},  while no compound with the stoichiometric ratio of \ce{Ti2Ir} has been reported. Hence, for both \ce{Ti4Rh2O} and \ce{Ti4Ir2O}, oxygen plays a crucial role in stabilizing the compounds in the $\eta$-carbide structure.

In figure \ref{fig:figure4}(c), we show temperature-dependent resistivity measurements of \ce{Ti2Co} and for comparison of the oxygen filled version \ce{Ti4Co2O} in the temperature range between $T$ = 400 mK and 300 K. We find that the resistivity of \ce{Ti2Co} slightly increases with decreasing temperature, characteristic for a semi-metallic behavior. While, the resistivity of the \ce{Ti4Co2O} with oxygen in the void positions is clearly metallic with a decreasing resistivity with decreasing temperature. The room-temperature resistivities of \ce{Ti2Co} and \ce{Ti4Co2O} are very similar, with ${\rho(300K)}$ = 0.077  m$\Omega$ cm and ${\rho(300K)}$ = 0.066  m$\Omega$ cm, respectively. While the metallic \ce{Ti4Co2O} clearly undergoes a transition to a superconducting state at a critical temperature $T_{\rm c}$ of 2.7 K, the semimetallic \ce{Ti2Co} shows a slight indication for a superconducting transition below 750 mK, as shown in figure \ref{fig:figure4}(d). The transition for \ce{Ti2Co} is very broad, and the resistivity at the lowest measured temperature here ($T_{\rm min}$ = 400 mK) is still above a zero resistance state. Hence, a possible complete transition to a superconducting state is not observed due to the temperature limitation of our measurement device. Based on the commonly used 50\% resistivity criteria, we estimate the superconducting temperature of \ce{Ti2Co} to be 0.5 K. This finding is in contrast to an earlier report showing superconductivity at higher temperatures\cite{ku1984new}. It is likely that the samples in the previous reports actually had oxygen incorporated in the interstitial positions, making them superconducting at higher temperatures. 

In figure \ref{fig:figure4}(e), the specific heat data in the normal state of \ce{Ti2Co} and \ce{Ti4Co2O} single crystals is shown. The value for the Sommerfeld parameter $\gamma$ is found to be $\gamma =$ 5.84 mJ mol$^{-1}$ K$^{-2}$ for \ce{Ti2Co} and $\gamma =$ 39.98 mJ mol$^{-1}$ K$^{-2}$ for \ce{Ti4Co2O}, respectively. As $\gamma$ is proportional to the electronic density of states at the Fermi-level $D$($E_{\rm F}$), the nearly 4-fold increase from \ce{Ti2Co} to \ce{Ti4Co2O} upon the addition of oxygen explains the semimetallic to metallic change in the electronic properties observed in the resistivities. Hence, the oxygen addition into the voids of \ce{Ti2Co} corresponds to an increase in electrons at the Fermi-level, leading to a metallic sample that eventually can become superconducting at a higher critical temperature. 

The phononic coefficient $\beta$ is similar for both compounds, with a value of $\beta =$ 0.096 mJ mol$^{-1}$ K$^{-4}$ for \ce{Ti2Co} and $\beta$ = 0.24 mJ mol$^{-1}$ K$^{-4}$ for \ce{Ti4Co2O}. The Debye temperature $\Theta_{\rm D}$ is calculated to be $\Theta_{\rm D} =$ 392 K for \ce{Ti2Co}, and $\Theta_{\rm D} =$ 386 K for \ce{Ti4Co2O}, indicating -- as it would be expected -- very similar phononic properties of the two compounds. We find both \ce{Ti2Co} and \ce{Ti4Co2O} to be Pauli paramagnets in the normal state (see Supporting Information).

\begin{figure}
 \centering
\includegraphics[width= \textwidth]{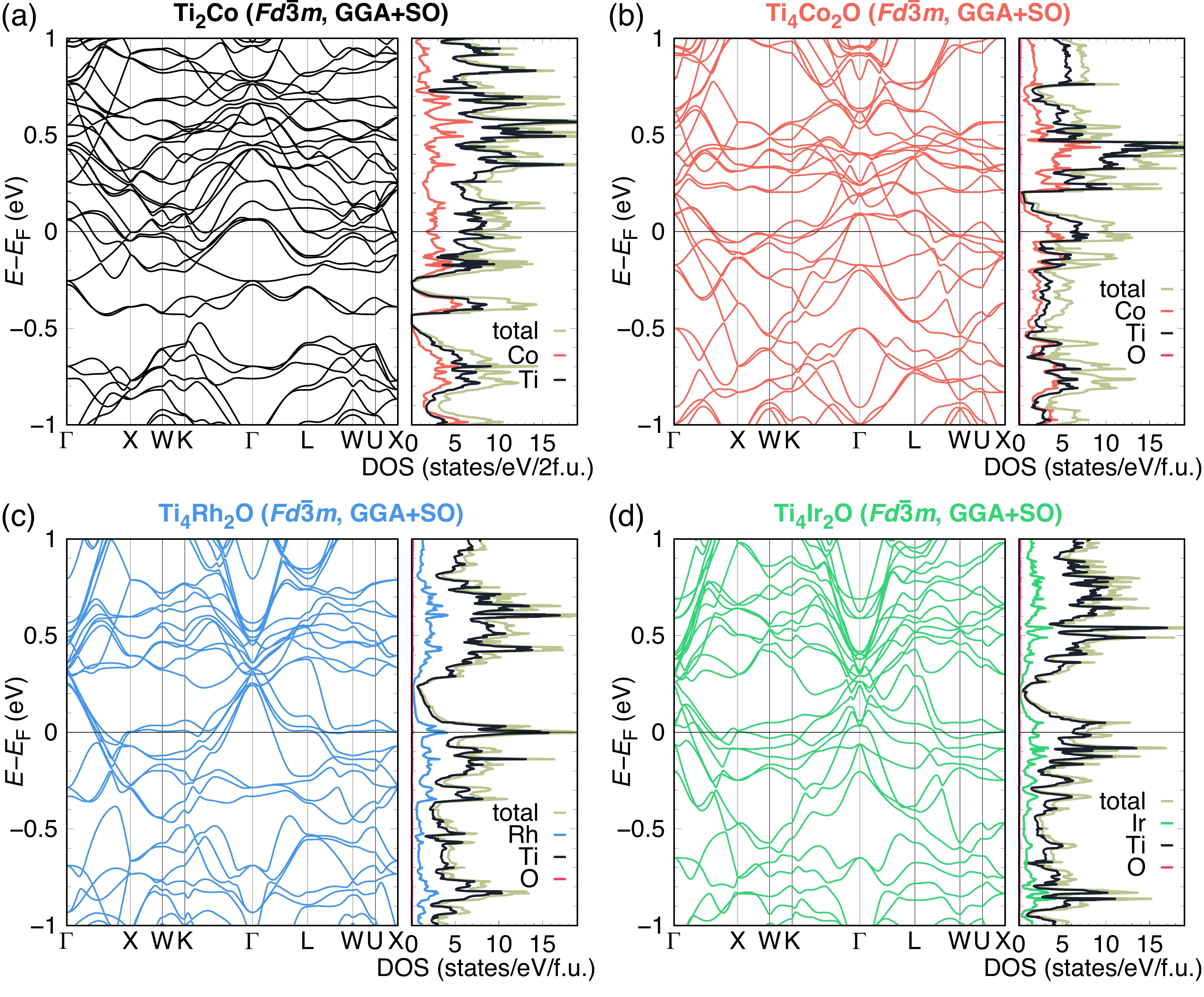}
\caption{Band structures and densities of states for (a) \ce{Ti2Co}, (b) \ce{Ti4Co2O}, (c) \ce{Ti4Rh2O}, and (d) \ce{Ti4Ir2O} respectively. The fully relativistic FPLO calculations with GGA functional include the effects of spin-orbit coupling. }
\label{fig:figure5}
\end{figure}

\section{Electronic structure of \ce{Ti4M2O} (M=Co, Rh, Ir)}

We use DFT calculations to better understand the electronic structure of the four materials \ce{Ti2Co} and \ce{Ti4$M$2O} ($M$=Co, Rh, Ir). Figure~\ref{fig:figure5} shows the fully relativistic GGA band structures and densities of states, labeled as GGA+SO. Overall bandwidth of Ti $3d$ states and Co $3d$, Rh $4d$ or Ir $5d$ states, respectively, is about 8\,eV. Filled oxygen $2p$ states are below -6\,eV and are not shown here. As we are interested in superconductivity, we focus on the states in a small window near the Fermi level $E_{\rm F}$. For easy comparison with the three other compounds, we normalize \ce{Ti2Co} DOS to two formula units. We find that even in the two Co-containing materials, density of states at the Fermi level $N(E_{\rm F})$ is affected by SO coupling. At least at the GGA+SO level, DFT does not stabilize magnetic moments on the transition metals. Our calculations yield a density of states at the Fermi-level of $D(E_{\rm F})$ = 7.99 states/eV/2f.u. in \ce{Ti2Co} and $D(E_{\rm F})$ = 10.2 states/eV/f.u. in \ce{Ti4Co2O}; thus, plain GGA+SO calculations can explain that $N(E_{\rm F})$ is smaller in the former but not that it is reduced by a large factor as seen from our specific heat measurements; this merits further investigation.

The difference between \ce{Ti2Co} and \ce{Ti4Co2O} is that in the latter, the Fermi level is placed 0.2\,eV lower within the $d$ states due to the $2e$ per formula unit transferred to O$^{2-}$, evidencing the strong electron-accepting character of the incorporated oxygen atoms into the void position. The binding of Ti2 to O also leads to substantial changes in the band dispersion so that the similarity between \ce{Ti2Co} and \ce{Ti4Co2O} band structures can only be recognized in special points like $\Gamma$, for example. The small shift of the Fermi level leads to a 28\% increase in $D(E_{\rm F})$ in \ce{Ti4Co2O} compared to \ce{Ti2Co}, providing a possible explanation why the former is superconducting, but the latter is not. On the other hand, DFT does not provide a simple argument why the critical temperature $T_{\rm c}$ of \ce{Ti4Ir2O} is nearly twice that of \ce{Ti4Rh2O}. The strong spin-orbit coupling in Ir leads to substantial band splittings and therefore a significantly lower density of states at the Fermi level, in agreement with the experimental findings. As the Fermi surface of \ce{Ti4Ir2O} is clearly more complex than that of \ce{Ti4Rh2O} due to spin orbit coupling, it will be interesting to investigate in a future study if better Fermi surface nesting enhances the electron-phonon coupling constant in \ce{Ti4Ir2O} and is thus responsible for the higher $T_{\rm c}$ and possibly the Pauli limit violation.

\subsection{Conclusion}

We have successfully synthesized $\eta$-carbide type suboxide superconductors \ce{Ti4M2O} (M=Co, Rh, Ir) and measured their superconducting properties. Magnetic susceptibility, electrical resistivity and specific heat capacity measurements showed \ce{Ti4M2O} (M=Co, Rh, Ir) are type-II superconductors with transition temperatures of $T_{\rm c}$ = 2.7, 2.8, and 5.4 K, respectively. The normalized specific heat jumps, $\Delta C/\gamma T_{\rm c}$, are measured to be 1.65, 1.28, and 1.80 for \ce{Ti4Co2O}, \ce{Ti4Rh2O}, and \ce{Ti4Ir2O}, evidencing the bulk nature of the superconductivity in these materials. 

Temperature- and field-dependent resistivity and specific heat measurements showed that all three compounds display high upper critical fields. The upper critical fields of \ce{Ti4Co2O} and \ce{Ti4Ir2O} are even by far exceeding the weak-coupling Pauli limit, indicating the unusual nature of the superconducting state in these compounds. Especially noteworthy is the superconductor \ce{Ti4Ir2O}, for which we find an upper critical field of $\mu_0H_{\rm c2}$(0) = 16.06 T, while the maximal expected upper critical field, i.e. the weak-coupling Pauli limit is $\mu_0H_{\rm Pauli}$ = 9.86 T. The observation of this violation in a cubic centrosymmetric material, in the absence of any magnetic interactions is unexpected. The reason for the absence of the weak-coupling Pauli limit violation in \ce{Ti4Rh2O} is not obvious. The upper critical fields of \ce{Ti4Rh2O} are, however, in very good agreement with the upper critical field observed in the $\eta$-carbide \ce{Z4Rh2O}, where an $\mu_0 H_{\rm c2}$(0) very close to the weak-coupling Pauli limit was observed, too.\cite{ma2019superconductivity} If the Pauli limit violation in \ce{Ti4M2O} was caused by strong-spin orbit coupling (the most commonly expected cause), then the effect would be expected to enhance across the group-9 transition metal series. The difference might rather be caused by the more complex Fermi surface. Furthermore, it is worth emphasizing that the specific heat jumps $\Delta C/\gamma T_{\rm c}$ of \ce{Ti4Co2O} and \ce{Ti4Ir2O} are clearly above the BCS value of 1.43, which might hint towards a strong coupling scenario for these two superconductors.

However, \ce{Ti4Co2O} and \ce{Ti4Ir2O} are also two additional examples for the observation of this violation after the recent report on \ce{Nb4Rh2C_{1-\delta}} in the same structure type. This indicates that the overall observed large upper-critical field and the weak-coupling Pauli limit violation are common properties that run in this structural family of compounds.

Our results show that in the class of the $\eta$-carbides and related structures, new superconducting materials can be discovered by using chemical design principles. Furthermore, the remarkably high upper critical fields discovered in these materials here and also in reference \citenum{ma2021superconductivity} may spark significant interest. Future investigations into the nature and origin of this effect -- which eventually is most relevant for the generation of next generation superconducting magnets -- can be expected.

\begin{acknowledgement}

This work was supported by the Swiss National Science Foundation under Grant No. PCEFP2\_194183. R.L. acknowledges a personal grant by the Forschungskredit of the University of Zurich. The work at Gdansk Tech. was supported by the National Science Center (Poland), grant number: UMO-2017/27/B/ST5/03044.

\end{acknowledgement}

\begin{suppinfo}

Photographs of  \ce{Ti2Co}, \ce{Ti4Co2O},  \ce{Ti4Rh2O}, and \ce{Ti4Ir2O} single crystals, crystallographic data for single-crystals of \ce{Ti4Rh2O} and \ce{Ti4Ir2O},  heat capacities of \ce{Ti2Co} and \ce{Ti4Co2O} single crystal under varying temperatures,  normal state magnetization of \ce{Ti2Co} and \ce{Ti4Co2O}, resistivities of \ce{Ti4Rh2O} and \ce{Ti4Ir2O} between 2 and 300 K, normal state magnetization of \ce{Ti4Rh2O} and \ce{Ti4Ir2O}, heat capacities of \ce{Ti4Rh2O} and \ce{Ti4Ir2O} under varying temperatures and magnetic fields, lower upper critical field measurements and fitting of \ce{Ti4Ir2O}, calculated band structures and densities of states for \ce{Ti2Co},  \ce{Ti4Co2O}, \ce{Ti4Rh2O}, and \ce{Ti4Ir2O} respectively. 

\end{suppinfo}

\bibliography{Ti4M2O}

\end{document}